\documentstyle[prl,aps,floats,psfig]{revtex} 
\begin{document}
\twocolumn[
\hsize\textwidth\columnwidth\hsize\csname@twocolumnfalse\endcsname

\title{Spin injection through the depletion layer: a theory of
spin-polarized {\it p-n} junctions and solar cells}

\author{Igor \v{Z}uti\'{c}$^1$, Jaroslav Fabian$^{1,2}$, and 
S. Das Sarma$^{1}$} 
\address{$^1$Department of Physics, University of Maryland at  College
Park, College Park, Maryland 20742-4111, USA\\
$^2$Max-Planck Institute for the Physics of Complex Systems, 
N\"{o}thnitzer Str. 38, D-01187 Dresden, Germany}

\maketitle

\begin{abstract}
A drift-diffusion model for spin-charge transport in spin-polarized {\it p-n}
junctions is developed and solved numerically for a realistic set of material
parameters based on GaAs. It is demonstrated that spin polarization can be
injected through the depletion layer by both minority and majority carriers,
making all-semiconductor devices 
such as
spin-polarized solar cells and 
bipolar transistors
feasible. Spin-polarized {\it p-n} junctions allow for
spin-polarized current generation, spin amplification, voltage control of
spin polarization, and a significant extension of spin diffusion range.
\end{abstract}
]

Spintronics\cite{prinz98} has played an important role in defining novel 
applications that are either not feasible or ineffective with
traditional semiconductor electronics. Spintronic devices
have found their niche in industries 
for magnetic read heads and nonvolatile memory cells.
Here we propose and demonstrate a new scheme for spintronics,  
a spin-polarized {\it p-n} junction, which amplifies spin density, 
significantly extends the range of spin 
diffusion, electronically tailors spin 
polarization, and, in combination with light as a spin-polarized 
solar sell, generates spin-polarized currents with tunable spin 
polarization. We prove these concepts
by solving drift-diffusion equations for a realistic device model
based on GaAs, which demonstrates that spin 
polarization can be injected through the depletion layer.
Possibility of injecting spin polarization through a transistor
is also discussed.

The electrical injection of spin-polarized carriers within all-semiconductor
structures (from a magnetic into a nonmagnetic semiconductor)
was recently realized experimentally 
\cite{fiederling99} (the scheme proposed in \cite{oestreich99}).
Optical injection of spin-polarized carriers (both minority
\cite{orientation84,zerrouati88} and majority
\cite{orientation84,awschalom97}) has been known for some time. 
In addition, the relatively
long spin diffusion lengths\cite{awschalom97,hagele98},
coherent spin transport across semiconductor
interfaces, 
a successful fabrication of a magnetic/nonmagnetic {\it p-n} 
junction\cite{ohno00} based on the novel (Ga,Mn)As 
material\cite{ohno98},
and the recent demonstration of a gate-voltage control of magnetization
in (In,Mn)As\cite{ohno00a}, make semiconductors promising materials 
for spintronic applications
\cite{dassarma00}. 

In this paper we investigate the
spin-charge transport in semiconductors under the conditions
of inhomogeneous bipolar doping (Flatte and
Vignale\cite{flatte01} have recently made an interesting proposal
for spin diodes and transistors in unipolar semiconductor
heterostructures--a very different case from ours); we 
are not concerned with spin injection {\it per se}. Our 
model device is a spin-polarized {\it p-n} junction with spin
polarization induced (either optically--in which case we get 
a spin-polarized solar cell--or electronically) to minority
or majority carriers. By studying spin-charge transport numerically
across the depletion layer, we observe novel phenomena, all 
resulting from the fact that spin polarization is transferred
(what we call injected) through the depletion layer.

We first introduce a drift-diffusion model for spin-charge
transport in  an inhomogeneously
doped semiconductor illuminated with circularly polarized light 
(and later solve the model for GaAs).
In addition to the approximations used in deriving the usual
(unpolarized) equations \cite{ashcroft76}, we assume that all the
dopants are fully ionized, the carrier populations
nondegenerate and varying only in one ($x$) direction. Further,
we assume the spin polarization to be carried solely by conduction
electrons (that is, consider holes unpolarized), as is the case
of III-V semiconductors like GaAs\cite{orientation84}, best
candidates for photo-spintronics.

The following parameters describe our model: acceptor (donor)
densities $N_A$ ($N_D$); electron (hole) densities $n$ ($p$),
in equilibrium $n_0$ ($p_0$); intrinsic carrier density $n_i$
($n_i^2=n_0p_0$); electron (hole) number current densities
$J_n$ ($J_p$); electron (hole) mobilities and diffusivities
$\mu_n$ ($\mu_p$) and $D_n$ ($D_p$); bipolar photoexcitation
rate $G$; intrinsic generation-recombination rate constant $w$
and spin-relaxation time $T_1$. Electron parameters $n$, $J_n$, and
$G$ will carry spin index $\lambda$ ($\lambda=1,\uparrow$ for spin up
and $\lambda=-1,\downarrow$ for spin down):
$n_{\uparrow}+n_{\downarrow}=n$, $J_{n\uparrow}+J_{n\downarrow}=J_n$,
and $G_{\uparrow}+G_{\downarrow}=G$. We also define spin-related
quantities: spin density
$s=n_{\uparrow}-n_{\downarrow}$, spin polarization
$\alpha=s/n$, and spin photoexcitation rate $G^s=G_\uparrow-G_\downarrow$.

The time evolution and spatial distribution of carrier and spin densities is
described by three sets of equations.
(i) Poisson's equation $d^2\phi/dx^2=-\rho/\epsilon$,
where $\phi$ is the electrostatic potential (related to the electric field
$E=-d\phi/dx$ in the $x$ direction),
$\epsilon$ is the sample dielectric permittivity,
and $\rho=e(N_D-N_A-n+p)$ is the
local charge density with the elementary charge $e$.
(ii) The linear response equations for the particle currents,
$J_{n\lambda}=-\mu_n n_{\lambda}E-D_n(dn_\lambda/dx)$ and
$J_{p}=+\mu_p pE-D_p (dp/dx)$.
The mobilities and conductivities are connected via Einstein's
relation $eD=k_BT\mu$; $k_B$ is the Boltzmann constant, 
and $T$ the absolute temperature.
(iii) The continuity
equations
\begin{eqnarray}\label{eq:recombinationa}
\frac{dn_\lambda}{dt}+\frac{dJ_{n\lambda}}{dx}&=&-w(n_\lambda p -n_0p_0/2)-
\frac{n_\lambda-n_{-\lambda}}{2T_1}+G_{\lambda},\\
\label{eq:recombinationb}
\frac{dp}{dt}+\frac{dJ_{p}}{dx}&=&-w(n p -n_0p_0)+G,
\end{eqnarray}
expressing
the particle generation and recombination as well as spin
relaxation. Equation sets (i-iii), together with
appropriate boundary conditions drawn from the actual physical situation,
fully determine the steady state ($dn_\lambda/dt=dp/dt=0$) 
distribution of carrier and spin densities.

In a homogeneous case and steady state Eqs.~\ref{eq:recombinationa}
and \ref{eq:recombinationb} become
$w(np-n_0p_0)=G$, and $wsp+s/T_1=G^s$.
Note that the rate at which electron spin relaxes is not $1/T_1$, but rather
$wp+1/T_1$, reflecting the disappearance of  spin by recombination (spin is
effectively transferred to holes which then lose it). This is more pronounced
in $p$-doped samples. Let us see if our equations recover what is already
known about spin polarization in semiconductors. For polarization we get
$\alpha=\alpha_0(1+n_0p_0/np)/(1+1/T_1wp)$,
where $\alpha_0=G^s/G$ is the polarization at time of pair creation.
Take a $p$-doped sample. Then $p\approx p_0$,
$n\ll n_0$, and
$\alpha=\alpha_0/(1+\tau/T_1)$, with electron lifetime $\tau=1/wp_0$,
expresses spin orientation\cite{orientation84}: if $\tau \ll T_1$,
which is usually the case,
$\alpha\approx \alpha_0$ and electron spins
are effectively oriented throughout the sample. In an $n$-doped
sample $wp\approx G/n_0$, and the spin polarization is
$\alpha_0/(1+n_0/GT_1)$. This is optical spin pumping\cite{orientation84}:
spin
polarization depends on the illumination intensity $G$ and is noticeable at
$G \gg n_0/T_1$. 

\begin{figure}
\centerline{\psfig{file=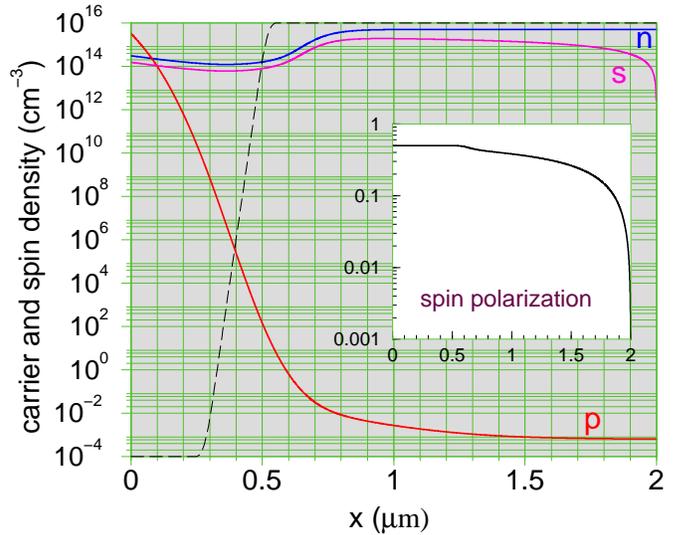,width=1\linewidth}}
\caption{Spin injection through the {\it p-n} junction. Electron
and hole densities $n$ and $p$ are
shown as a function of the distance $x$
from the illuminated surface ($x=0$).
The dashed line is the doping profile $N_D-N_A$ (scale not shown): it is
$p$-type with $N_A=3\times 10^{15}$ cm$^{-3}$ on the left and $n$-type with
$N_D=5\times 10^{15}$ cm$^{-3}$ on the right; the transition region is between
$0.25$ and $0.55$ $\mu m$.
Also plotted is spin density $s=n_\uparrow-n_\downarrow$ and spin
polarization $\alpha=n/s$ in the inset. The remarkable result that $\alpha$
extends well beyond the transition layer (and $s$ is amplified) demonstrates
both spin injection and spin
density amplification.
}
\label{fig:1}
\end{figure}

Our prototype model device is a $2$ $\mu$m long GaAs sample, doped with
$N_A=3\times 10^{15}$ cm$^{-3}$ acceptors on the left and with
$N_D=5\times 10^{15}$ cm$^{-3}$ donors on the right along the $x$-axis
(the doping profile is shown in Fig.~\ref{fig:1}).
The left surface of the sample, $x=0$, is illuminated by circularly
polarized light which creates electrons and holes with
photodensities $\delta n =\delta p= 3\times 10^{14}$ cm$^{-3}$,
and induces electronic spin polarization $\alpha_0=0.5$
(the value given by the band-structure symmetry\cite{orientation84})
at the surface. Since holes lose their spin orientation faster than
they gain it
(also a band structure effect\cite{orientation84}), their polarization
is effectively zero and need not be considered. The task is to
find the steady-state distribution of electron and hole densities
$n$ and $p$, as well as electron spin density
$s=n_\uparrow-n_\downarrow$ 
and polarization $\alpha=s/n$, as a function of $x$.

We solve the equation sets (i-iii) numerically, in the steady-state 
regime, 
with the boundary conditions: $p(0)=p_0(0)+\delta p$ (where 
$p_0(0)=0.5(N_A+(N_A^2+4n_i^2)^{0.5})$), $n(0)=p_0(0)/n_i^2+\delta n$, 
and $\alpha(0)=0.5$, reflecting bipolar photoexcitation and spin orientation
processes at the illuminated surface (there is no illumination in the 
bulk: $G=0$); $n(2)=n_0(2)$ (where $n_0(2)=0.5(N_D+(N_D^2+4n_i^2)^{0.5})$), 
$p(2)=n_0(2)/n_i^2$, and $\alpha(2)=0$, maintaining equilibrium at 
the right surface. 
For electrostatic potential $\phi$ we use 
$\phi(0)=-(k_BT/e)\ln(p_0(0)/n_i)$ and  
$\phi(2)=(k_BT/e)\ln(n_0(2))/n_i)-V$, where $V$ is the applied bias, 
conventionally defined with respect to the dark 
built-in value\cite{fahrenbruch83}.
The room-temperature parameters of GaAs used in our model are 
\cite{fahrenbruch83}: 
 the intrinsic carrier density 
$n_i=1.8 \times 10^{6}$ cm$^{-3}$,  
electron and hole mobilities $\mu_n=10\mu_p=4000$ 
${\rm cm}^2\cdot{\rm V}^{-1}\cdot{\rm s}^{-1}$,
electron and hole diffusivities (from Einstein's relation) 
$D_n=10D_p=103.6$ ${\rm cm}^2\cdot{\rm s}^{-1}$, and dielectric 
permittivity 
$\epsilon=13.1 \epsilon_0$ (where $\epsilon_0$ is the permittivity of 
free space). As for the generation-recombination rate $w$ (assumed to be
a constant independent of $n$ and $p$), we take the value of 
$(1/3)\times 10^{-5}$ ${\rm cm}^3\cdot{\rm s}^{-1}$. 
The electron lifetime in the 
$p$-region is then $\tau\approx 1/(wN_A)\approx 0.1$ ns. 
Finally, spin-relaxation time 
$T_1$ is taken to be $0.2$ ns, a reasonable value for GaAs
\cite{zerrouati88,awschalom97,fabian99}. 

We first discuss the case of $V=0$ (the applied reverse bias from an 
external battery cancels the forward voltage due to
photoexcitation), which is in Fig.~\ref{fig:1}. 
Electronic density starts off with the value of $n\approx \delta 
n$ at $x=0$, decreases somewhat in the 
depletion layer, then rises 
by more than a decade to reach its equilibrium value of $n_0\approx N_D$
in the $n$ region. The hole density sharply decreases 
from approximately $N_A$ through the depletion layer until it becomes  
$\sim n_i^2/N_D$, the equilibrium value on the $n$ side.
Spin density $s$ is $0.5n$ at the 
illuminated surface, and essentially follows the 
spatial dependence of $n$ through the depletion layer, but once in the $n$
region it decays towards zero\cite{boundary}. The surprising
feature is the increase of $s$ through the depletion layer. In effect,
the magnetization 
of the sample increases by more than an order of magnitude as a result of  
spin 
injection. The polarization $\alpha$ (Fig.~\ref{fig:1}, inset) starts at 0.5 
at the illuminated surface, stays almost constant through the transition 
region, then decreases to zero at the right boundary\cite{boundary}.
We checked that  these results are robust against changes in $T_1$
and $\tau$ by up to two decades, as long as $T_1\agt \tau$ (so that 
appreciable spin polarization can be induced in the $p$ region in
the first place).

Spin polarization clearly survives the depletion layer. 
This is not an ordinary spin injection in which certain
number of spin-polarized electrons tunnel through a contact and the spin
density is equal on both sides. What we have, rather, is a spin pumping 
(leading to spin density 
 amplification) mechanism. Indeed, in optical spin pumping
\cite{orientation84} circularly polarized light creates spin 
polarization of majority carriers (electrons in $n$ region) by intensive
illumination which generates spins at a faster rate than
$1/T_1$. Here we illuminate the $p$ (not $n$)  region, so we do not 
have optical spin pumping. The physics
is the following: light induces spin polarization of minority carriers
(electrons in the $p$ region) through optical spin orientation 
(see Ref.~\onlinecite{orientation84}), which diffuse towards 
the transition region
where they are swiftly pushed by the built-in field into their native
$n$ region. What we have is  {\it spin pumping through 
the minority channel}: spin-polarized minority electrons bring spin into 
the $n$ region faster than the spin there relaxes or diffuses away.

In a sense, minority electrons play the role of circularly polarized 
light. The evidence for that is in Fig.~\ref{fig:1}. 
In the $n$ region, spin imbalance is present even though the carrier
populations are well relaxed in equilibrium. As a result, spin diffusion
is controlled by the majority diffusivity constant $D_n$\cite{orientation84}
rather than the minority constant 
$D_p$, as would be  the case of carrier 
diffusion. Spin decay beyond the depletion layer
is therefore described by the single exponential $s\sim \exp(-x/L_s)$,
where $L^n_s=(D_nT_1)^{0.5}\approx 1.4$ $\mu$m is the spin diffusion
length of electrons in the $n$ region. In the $p$-region spin diffusion
length is $L^p_s=(D_p\tau_s)^{0.5}$;
electrons diffuse bipolarly with hole diffusivity $D_p$
for a time $\tau_s=\tau T_1/(\tau+T_1)\approx0.067$ ns,
so that $L^p_s\approx 0.26$ $\mu$m is
of the order of 
the electron diffusion length in the $p$-region. Since
$L^n_s/L^p_s=(D_n/D_p)^{0.5}\times(1+T_1/\tau)^{0.5}$ ($\approx 5$ in our
model), the effective
range of spin diffusion is extended far beyond the expected value.
The depletion layer acts to extend the range of spin diffusion, with 
the effect most pronounced for $T_1\gg \tau$, which is typically 
the case (especially at lower temperatures).

Our model scheme above constitutes a solar cell, as
light illuminates the surface within the electron diffusion length
from the depletion layer. As 
the electron diffusion length almost coincides
with $L^p_s$, electrons arrive at the transition region spin polarized.
The built-in electric field then  
sweeps the electrons in
 the $n$-region (and holes
back into $p$). Because of the spin amplification in the $n$-region,
the resulting electrical current is spin polarized and can be  used
for spintronic applications (in a combination with ferromagnetic
semiconductors or metals). 
Our spin-polarized
solar cell has the usual I-V characteristics, with 1.03 V open-circuit
voltage.

\begin{figure}
\centerline{\psfig{file=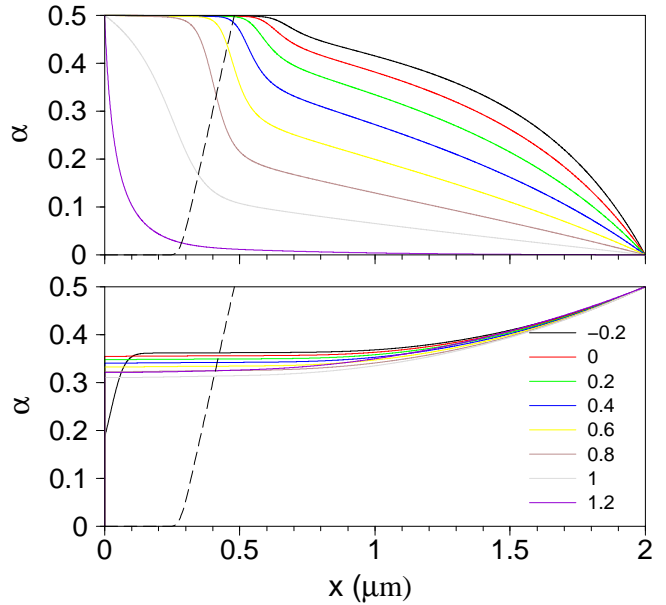,width=1\linewidth}}
\caption{Calculated spin polarization ($\alpha$)
profiles for different voltages.
The upper figure is for the solar cell. The lower figure is for an injection
of spin by a ferromagnetic electrode at $x=2\mu {\rm m}$. In both cases
$\alpha$ extends beyond the transition region, while the strong
dependence of $\alpha$ on $V$ is seen only in the solar cell. The dashed
line in both figures is the doping profile.
}
\label{fig:2}
\end{figure}

\begin{figure}
\centerline{\psfig{file=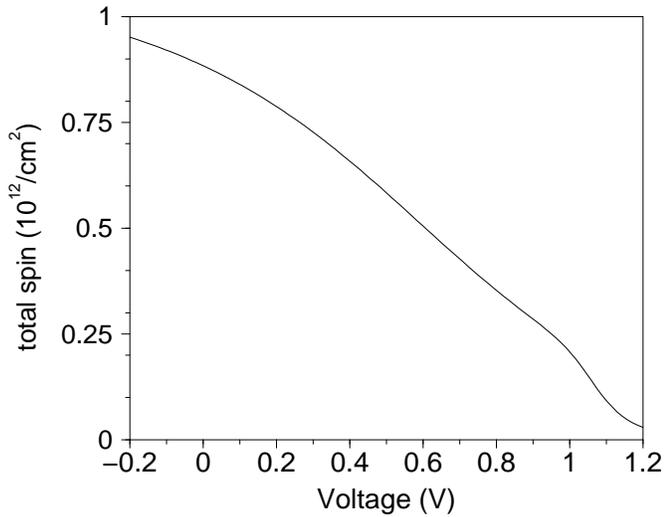,width=1\linewidth}}
\caption{Total spin in the solar cell as a function of applied
voltage.
}
\label{fig:3}
\end{figure}

How does spin polarization 
$\alpha(x)$ change with the applied bias 
 $V$? The calculated profiles 
for our model device are in Fig.~\ref{fig:2} (top). 
There is a strong monotonic decrease of $\alpha$ 
with increasing $V$. To quantify this dependence
we consider the change of the total spin accumulated in the 
cell (integral of the spin density 
$s$ from $x=0$ to 2); the spin accumulates almost
exclusively in the $n$ region (where electron density
$n$ is large). The result
is in Fig.~\ref{fig:3}. Total spin changes by almost 20 times when 
increasing $V$ from -0.2 V to 1.2 V. 
By (loose) analogy with junction capacitance, we call this effect 
spin capacitance.  The spin accumulation
in the $p$ region essentially follows the nonequilibrium density
of electrons (charges) there,  but in the $n$ region, where spin 
diffusion length is much greater than carrier diffusion length, 
nonequilibrium spin accumulates to a much greater
distance; spin capacitance is not trivially connected with
nonequilibrium charges. Clearly, after switching off the light 
the spin is lost, so to ``store'' spins in the solar cell one 
needs to supply energy (the fact that $T_1\agt \tau$ will lead to
special transient effects related to spin recovery).
The reason why $\alpha$ depends so strongly on $V$ 
is that  $V$ changes
the extent of the depletion layer: as $V$ increases, the width of the
depletion layer decreases\cite{ashcroft76}. Since we are illuminating 
the same point $x=0$ in all cases, the amount of spin polarization 
that reaches the depletion layer decreases as the width of the layer
decreases (so that the distance from the surface to the center of the
layer increases). At large forward voltages (say, $V=1.2$ V, in 
Fig.~\ref{fig:2}), the injected spin essentially decays within $L^p_s$,
which is shorter than the distance from the illumination surface to
the 
depletion layer. Another effect affecting the dependence of $\alpha$
on $V$ is the value of the built-in field, which also decreases with
increasing $V$. This electronic control of spin polarization could
be measured by observing luminescence of electrons in the $n$ region
(by, say, forcing them to recombine with holes in heterostructures
attached to the right surface). The above effects are
not limited to optical spin injection. The only requirement
is that there is an electronic spin and carrier imbalance (which can be 
also created electronically) in the {\it p} region.

Finally, consider the case of spin polarization injected in the majority
(here $n$) region (say, by ferromagnetic electrodes). 
There is no light illuminating the sample, and all 
the parameters and 
boundary conditions remain as above, except that at $x=0$ all the
charge  densities
are the equilibrium ones, and $\alpha(2)=0.5$ (a quite favorable case). 
The calculated $\alpha(x)$ for different $V$ are in Fig.~\ref{fig:2} (bottom).
The polarization, as before, is injected through the depletion layer
(into minority electrons), but now $V$ does not affect $\alpha(x)$ that much.
Consider what would happen if an $n$ region was now attached to
the sample from the left, that is, we have an {\it n-p-n} transistor.
The spin-polarization in the base ($p$) would be injected
and amplified in the $n$-collector (by the spin-pumping from the
minority channel we introduced above). We conclude that spin polarization 
can be injected all the way through a transistor, from emitter into collector. 

We thank Paul Crowell for useful discussions. This work was supported by 
DARPA and the U.S. O.N.R.


\begin{references}
\bibitem{boundary} The spin decay beyond 1.5 $\mu$m is not a bulk effect.
It is caused by our boundary condition $\alpha(2)=0$ which is quite severe
(it may physically correspond to a fast spin relaxation at that boundary
caused by magnetic impurities), although it does not affect the important
region around the depletion layer.
\end{references}
\end{document}